\documentclass[]{spie}  %>>> use for US letter paper
\usepackage[]{graphicx}
\usepackage{url}
\usepackage[linkcolor=blue]{hyperref}
\usepackage[lmargin=2.22cm,rmargin=2.22cm,tmargin=2.54cm,bmargin=3.17cm]{geometry}
\usepackage{mathptmx}
\usepackage{amsmath}
\usepackage{amssymb}
\usepackage{amsfonts}
\usepackage{times}
\usepackage{amsmath}
\usepackage{amssymb}
\usepackage{amsfonts}
\usepackage{color}
\usepackage{enumerate}
\usepackage{caption} 
\usepackage{float}
\title{Single-Pass GPU-Raycasting for\\ Structured Adaptive Mesh
  Refinement Data}

\author{Ralf Kaehler\supit{a} and Tom Abel\supit{a}
\skiplinehalf
\supit{a}KIPAC/SLAC, 2575 Sand Hill Road, Menlo Park, USA
}

\authorinfo{Further author information: (Send correspondence to R.~Kaehler)\\
R.~Kaehler: E-mail: kaehler@slac.stanford.edu, Telephone: 1 650 926 2884\\ 
T.~Abel: E-mail: tabel@slac.stanford.edu, Telephone: 1 650 926 2421}
\begin{document} 
  \maketitle

%%%%%%%%%%%%%%%%%%%%%%%%%%%%%%%%%%%%%%%%%%%%%%%%%%%%%%%%%%%%% 
\begin{abstract}
Structured Adaptive Mesh Refinement (SAMR) is a popular numerical
technique to study processes with high spatial and temporal dynamic range. 
It reduces computational requirements by adapting the lattice on which the
underlying differential equations are solved to most efficiently
represent the solution. Particularly in
astrophysics and cosmology such simulations now can capture spatial
scales ten orders of magnitude apart and more.
The irregular locations and extensions of the refined regions in
the SAMR scheme and the fact that different resolution levels partially overlap, poses a
challenge for GPU-based direct volume rendering methods. 
kD-trees have proven to be advantageous to subdivide the data domain
into non-overlapping blocks of equally sized cells, optimal for the texture
units of current graphics hardware, but previous GPU-supported raycasting
approaches for SAMR data using this data structure required a separate
rendering pass for each node, preventing the application of
many advanced lighting schemes that require simultaneous access to more 
than one block of cells. In this paper we present the first single-pass 
GPU-raycasting algorithm for SAMR data that is based on a kD-tree.
The tree is efficiently encoded by a set of 3D-textures, which allows
to adaptively sample complete rays entirely on the GPU without any CPU interaction.
We discuss two different data storage strategies to access
the grid data on the GPU and apply them to several datasets
to prove the benefits of the proposed method.
\end{abstract}

%>>>> Include a list of keywords after the abstract 


%%%%%%%%%%%%%%%%%%%%%%%%%%%%%%%%%%%%%%%%%%%%%%%%%%%%%%%%%%%%%
\section{INTRODUCTION}
\label{Sec::Intro}  

Multi-scale phenomena are common in many areas of
research, in particular astrophysics and cosmology, fluid dynamics, and mechanical
engineering. An example is the formation of the first stellar
objects in the Universe, involving spatial scales that range from several 10,000 light years, 
representing the overall dynamics of the proto-galaxies, down to the
star forming regions in the order of a few light hours across~\cite{19589964,2002Sci...295...93A}.
Tackling such processes numerically is a challenging task, and naive
approaches using constant resolution fail due to their exorbitant
computational requirements.
Hence adaptive techniques are crucial for this type of problems, as they allow to 
locally adjust the spatio-temporal resolution to the 
features of the particular system. A popular adaptive approach for numerically solving
partial differential equations 
is {\sl Structured Adaptive Mesh Refinement (SAMR)}~\cite{berg:1984:olig}.
It combines the simplicity of structured grids with the benefits of 
local refinement by recursively overlaying regions of the
computational domain with patches of structured grids of increasing resolution.

Applying standard visualization techniques to AMR datasets has always
been challenging, partly due to the arbitrary extension and placement
of the subgrid patches, partly because of their sheer numbers. This
holds in particular for GPU-based volume rendering approaches, which
leverage the capabilities of texturing units of current graphics
hardware architectures.
These operate most efficiently on regular grids and therefore
a partitioning of the computational domain covered by the SAMR grids into
non-overlapping blocks of cells with the same resolution is crucial
for good performance.  Adaptive kD-tree have been proven to be
particularly suitable for this
tasks~\cite{Kaehler01AMRVolren,Kelly03remoteamr,weber:2001:amrvolren},
but previous GPU-based methods required a single rendering pass for
each of the resulting blocks, which inhibits the direct application of
many advanced shading and lighting effects that
need to simultaneously access data from more than one subgrid patch or
require non-standard blending equations.

In this paper we present a single-pass GPU-raycasting approach for AMR
data. It is based on an efficient kD-tree partition of the domain
that minimizes the number of generated nodes and can directly be
applied also to non-nested subgrids, which refine regions of more than
one coarse ``parent'' grid patch. We propose an efficient encoding of the resulting tree using a
set of 3D-textures, enabling the traversal of the tree and an
adaptive sampling of the data on the GPU, on a per-pixel basis
in the fragment shader, without any CPU interaction. 
We further discuss two different approaches to store the data
associated with the AMR grid patches: one
using a packing scheme to organize the patches in a larger 
{\sl memory pool} texture, the other employing {\sl NVIDIA's} {\sl Bindless
  Texture} extension~\cite{BindlessTexturesWebsite} for the {\sl OpenGL}-API.

The remainder of this paper is organized as follows. In
Section~\ref{Sec::RelatedWork} we discuss related work. We review the AMR
scheme in Section~\ref{Sec::AMR} and describe the new kD-tree generation strategy and its
encoding on the GPU in Section~\ref{Sec::RenderingAlgorithm}. The GPU-data access scheme
as well as the rendering algorithm will be discussed in
Section~\ref{Sec::GPUDataStorage}. We end with results and conclusions in
Section~\ref{Sec::Results} and~\ref{Sec::Conclusions}.
 
%%%%%%%%%%%%%%%%%%%%%%%%%%%%%%%%%%%%%%%%%%%%%%%%%%%%%%%%%%%%%
\section{RELATED WORK}
\label{Sec::RelatedWork} 
  
To the best of our knowledge the first CPU-based volume rendering method for
AMR data was proposed by Max~\cite{max:1993:sorting}. It employed a
back-to-front cell-sorting and cell-projection scheme.
Later the {\sl dual-mesh} approach~\cite{weber:2001:crackfreeiso}, for higher order interpolation of
``cell-centered'' AMR data without resampling, has been extended for
more general subgrid configurations and was used for a direct volume rendering approach~\cite{TVCG.2011.252}.
Further several parallel CPU-based volume rendering methods for
cluster architectures data have been presented~\cite{Weber:2003,0067-0049-192-1-9}.

The first GPU-supported volume rendering approach for AMR data
was presented by Weber~et~al.~\cite{weber:2001:amrvolren}. The authors 
applied their dual-mesh-``stitching'' scheme to implement a hardware-supported 
cell-projection algorithm rendering the faces of the resulting cells as
semi-transparent triangles. 
Kaehler~et~al. presented a 3D-texture-based volume rendering approach for large, sparse
datasets, that clusters non-transparent voxels into axis-aligned
blocks and encodes these as leaf nodes of AMR data structures.~\cite{Kaehler01sparseVolren} 
They also described a multi-resolution texture-based volume rendering
algorithm for AMR data~\cite{Kaehler01AMRVolren}.
Park~et~al.~\cite{park:vis02} presented a hierarchical splatting approach for
AMR data.  Kelley~et~al.~\cite{Kelly03remoteamr} describe a framework for
interactive, parallel volume rendering of remote AMR data,
that distributes subtrees of the AMR hierarchy on individual processors
and composes the images on a local rendering client. 

With the advent of programmable graphics hardware that
supports flexible shader programs, it became feasible to perform the
ray integration on a per-pixel basis at interactive frame rates~\cite{769948,1081482,stegmaier}.
In the latter approach the data is converted to a 3D texture and a 
fragment shader is executed for each pixel that is covered by the projected 
bounding box of the data volume. The ray is parameterized in 
texture coordinates and the ray-integral is computed in the
fragment shader.
GPU-raycasting is particularly attractive for adaptive grids, as it does not
suffer from the rendering artifacts inherent to slice-based
methods, which can lead to visible artifacts at the interfaces
between different resolution levels.
GPU-raycasting has been extended to SAMR data, using a kD-tree
that is traversed on the CPU and rendered node-by-node in 
separate rendering passes~\cite{vg06-kaehler}.

All previous approaches for single-pass
multi-resolution GPU-raycasting were based on regular data
structures, such as octrees or other partition strategies
using regularly shaped nodes~\cite{vg06-ljung,vg06-vollrath,Gobbetti:2008:SGR:1394332.1394367,CNLE09}. 
In principle also AMR data structures can be partitioned in blocks of
cells from the same resolution level using octrees. However, the
resulting tree is usually inefficient, in particular if higher order
interpolation is desired, because of the large number of
resulting nodes~\cite{Kaehler01sparseVolren}.
In contrast kD-trees allow to minimize the number of nodes by adaptively 
choosing the position of the spatial subdivision planes and have been
successfully applied to CPU-, and GPU-based volume rendering of AMR data~\cite{Kaehler01AMRVolren,Kelly03remoteamr,weber:2001:amrvolren,Marchesin:2009:HSV:1638611.1639161}.
In this paper we present the first single-pass GPU-raycasting approach for AMR
data based on a kD-partition of the data domain.

\section{STRUCTURED ADAPTIVE MESH REFINEMENT} \label{Sec::AMR}
In the {\sl Adaptive Mesh Refinement (AMR)}~\cite{berg:1984:olig} approach the computational domain is covered by a set of coarse, 
structured subgrids.The configuration of this set of coarse grids is usually fixed over time.
In a first step the solution of the (partial differential) equations
is computed on these coarse grids and local error estimators are utilized to detect cells 
that require higher resolution. These cells are clustered into a set of 
rectangular grid patches, usually called {\sl subgrids}, which
do not replace, but rather overlap the corresponding 
regions of the coarse base grids.
\begin{figure}[t]
  \centering
     \begin{minipage}[c]{0.33\textwidth}
         \resizebox{1\columnwidth}{!}{ \includegraphics{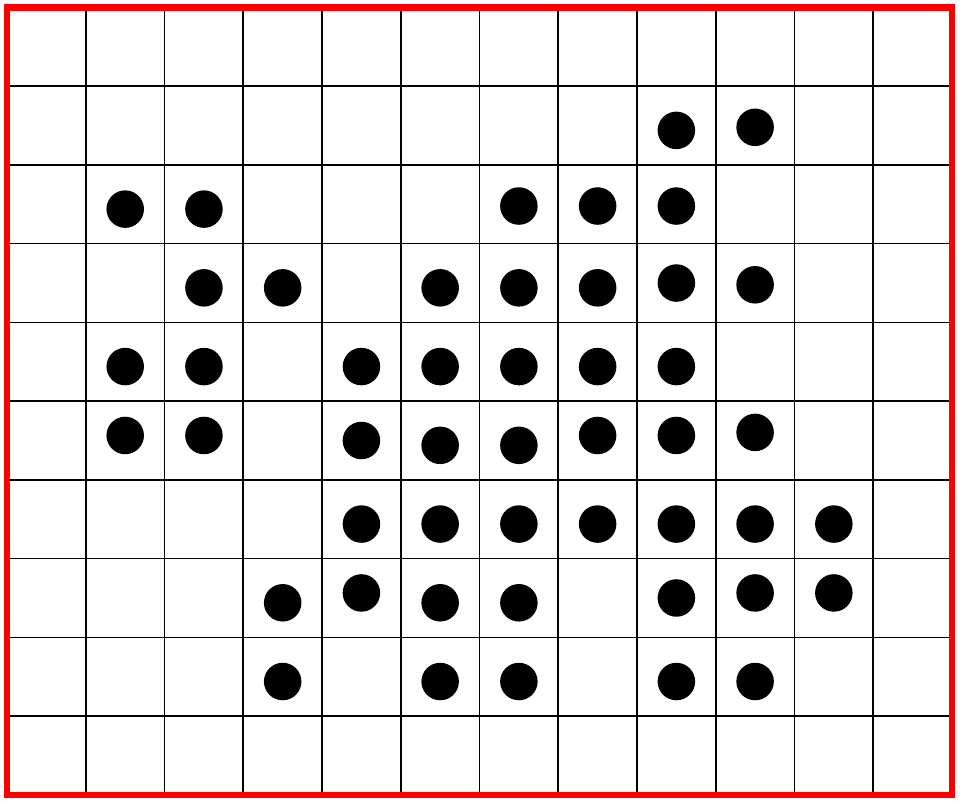} }
         \centerline{(a)}
     \end{minipage}
     \begin{minipage}[c]{0.33\textwidth}
        \resizebox{1\columnwidth}{!}{\includegraphics{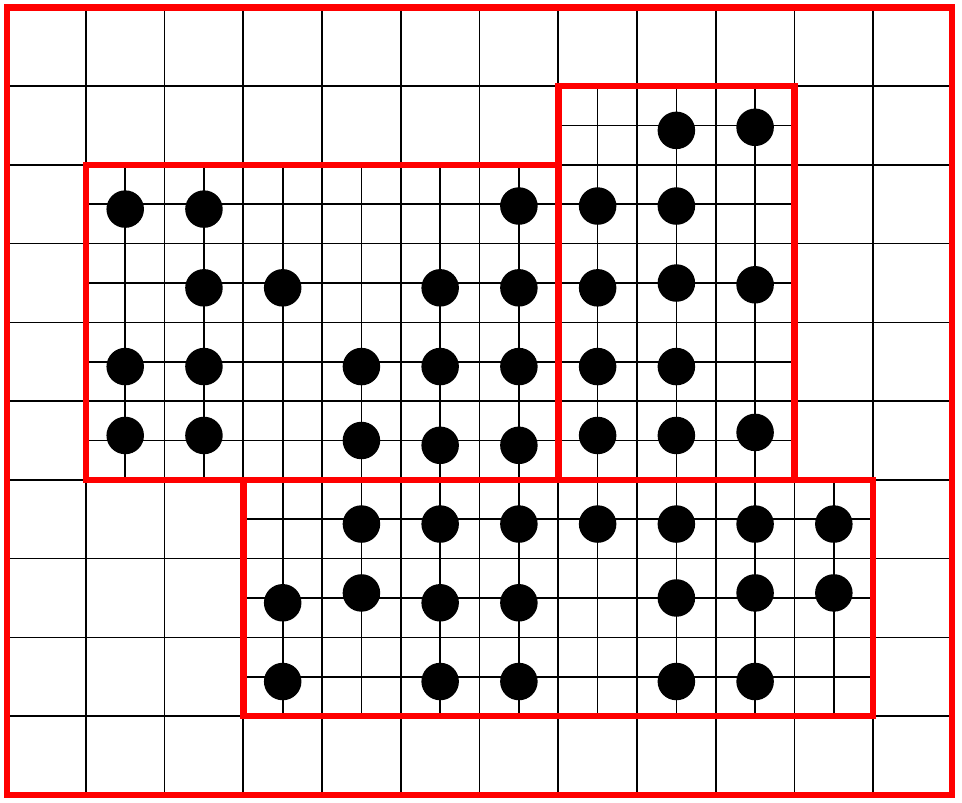} }
         \centerline{(b)}
     \end{minipage}
     \begin{minipage}[c]{0.33\textwidth}
        \resizebox{1\columnwidth}{!}{\includegraphics{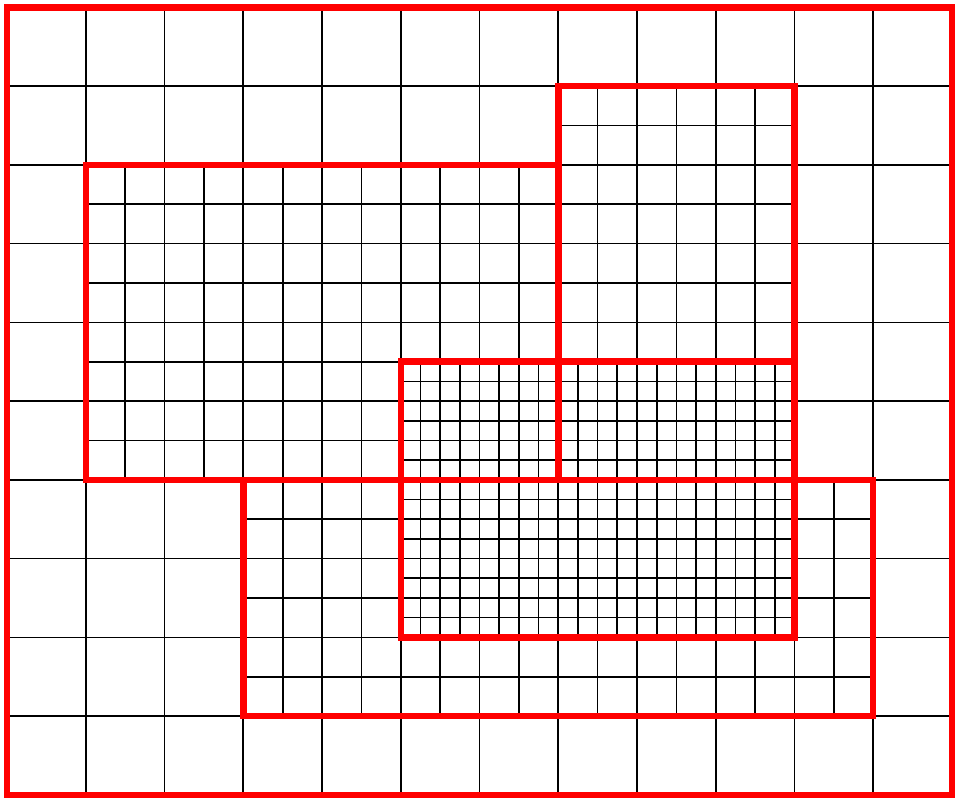} }
         \centerline{(c)}
     \end{minipage}
    \caption{Refinement process for AMR schemes: 
             Cells that require refinement are determined using local error 
             criteria~(a) and clustered into separate subgrids~(b), which
             cover the region at higher resolution. This process is 
             recursively continued~(c) until each region has sufficient resolution.}
    \label{Img:SpatialRefinementAMR}
\end{figure} 
The equations are solved on these higher resolved subgrids, and
the refinement procedure recursively continues until all cells have
sufficient resolution, giving rise to a hierarchy of nested refinement levels,
as shown in Figure~\ref{Img:SpatialRefinementAMR}.
A major advantage of AMR is that each subgrid can be viewed as a separate, 
independent structured grid with its separate storage space. This allows to process 
subgrids almost independently, and thus it is well-suited for parallel processing.
A popular variant of this general approach
 is {\sl Structured Adaptive Mesh Refinement
   (SAMR)}~\cite{ber:1989:col}, where in contrast to the original
 scheme, the subgrids are aligned with the major axes of the coordinate system.
In the following we will restrict the discussion to SAMR and just
refer to it as AMR. In the remainder of this section we will briefly introduce some notations that
are used in this paper.

\begin{figure} [h]
    \centering
    \resizebox{0.85\columnwidth}{!}{%
        \includegraphics{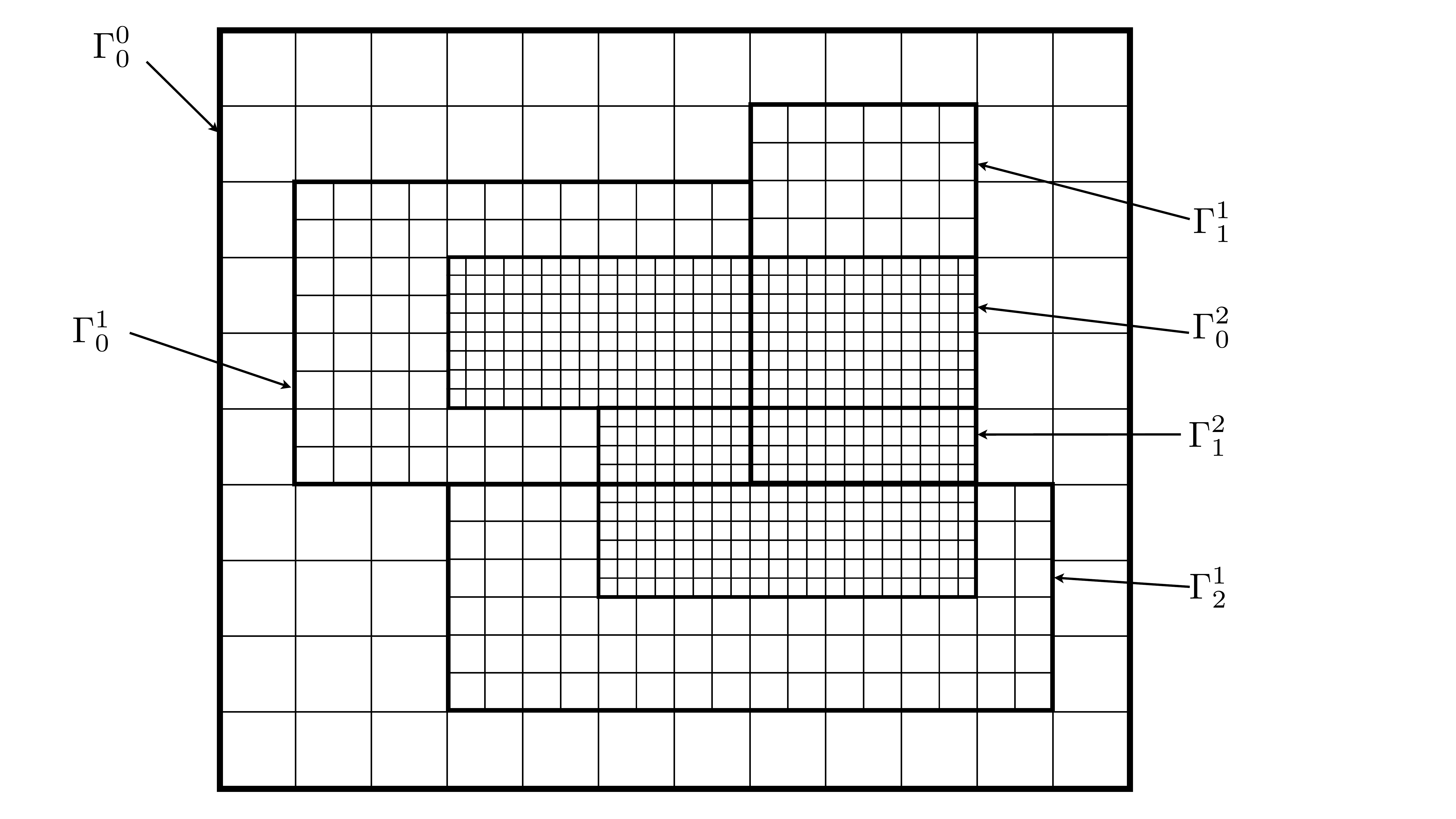}%
    }%
    \caption{Two-dimensional example of a hierarchy of structured AMR grids. In this
      case the root level is given by a single subgrid $\Gamma_0^0$,
      and is refined by three subgrids
             $\Gamma^1_0, \Gamma^1_1, \Gamma^1_2$, generating the
             first level of refinement. The first level is further refined by two subgrids 
             $\Gamma^2_0$ and $\Gamma^2_1$. The refinement
           level between  abutting cells can differ by more than one.}
    \label{Fig::GridStructure}
\end{figure}

Let ${\bf h}^{0} := (h^0_0,h^0_1,h^0_2)$  denote the mesh spacing
of the coarsest grids. The mesh spacings of
the finer grids are recursively defined by 
${\bf h}^{l} := (h^{l-1}_0/r,h^{l-1}_1/r,h^{l-1}_2/r)$ for $l>0$,
where the positive integer $r$ denotes the so-called {\sl refinement
  factor} and $l$ numbers the {\sl refinement level}, starting with
$0$ for the coarsest level. 
In principle the refinement factor can differ for each direction and each level, 
but in order to simplify the notation we assume that it is constant. 
In the AMR approach, each refined cell is overlaid by a set of 
$r^3$ cells of the next level of refinement. 
In the original AMR scheme~\cite{ber:1989:col} each refinement level 
was enclosed by at least one layer of cells from the next coarser level of 
resolution, such that adjacent cells differ by at most one level.
This constraint was later relaxed~\cite{Abel2002Sci,2004astro.ph..3044O}.
In the following we will call the set of coarsest subgrids the {\sl root} level and 
denote the $m$-th {\sl subgrid} of the refinement level~$l$ by
$\Gamma^{l}_{m}$, see Figure~\ref{Fig::GridStructure}.

\section{The Rendering Algorithm}
\label{Sec::RenderingAlgorithm}
The outline of our GPU-raycasting algorithm can be summarized as
follows:
\begin{itemize}
\item
First the hierarchy of nested refinement levels is decomposed into 
non-overlapping, axis-aligned blocks, each covering only cells from the 
same level of resolution. These are organized in an adaptive kD-tree
data structure, encoded using a set of integer-valued 3D textures. 
\item
  The data associated with the separate grids are either packed into
  a single 3D texture or accessed dynamically as individual textures
  using {\sl NVIDIA's Bindless Texture} extension for {\sl OpenGL}.
\item
 The textures are uploaded onto the GPU and the kD-tree is
  traversed in the fragment shader. For each pixel the intersections between the
  viewing ray and the nodes of the kD-tree are computed and the
  resulting ray-segments are processed in a front-to-back order.
  The color and opacity contribution of each segment is
  computed by adaptively sampling the corresponding textures, with a
  sampling distance based on the underlying
  level of refinement. The contributions are accumulated to
  yield the overall pixel color, which is written to the
  frame-buffer after all segments have been processed.
\end{itemize}
In the next subsections we will describe these steps in more detail.

\subsection{KD-TREE CONSTRUCTION} 
\label{Sec::kDTree}   
In order to leverage the capabilities of texturing units on current graphics
hardware, e.g.~for fast constant or tri-linear
interpolation on regular grids, it is beneficial to subdivide the data domain into
separate blocks which do not overlap and cover only cells from the same resolution level. 
Rendering a given hierarchy of separate subgrids directly would result in rendering 
artifacts, since in the AMR approach the subgrid patches on finer
levels do not replace but rather overlay regions of coarser levels, so 
refined regions of the data volume would be rendered multiple times. 

The root node of the kD-tree is defined by the enclosing bounding box ${B}$
of all subgrids $\Gamma^0_i$ on the root level of the AMR 
hierarchy. ${B}$ is recursively subdivided by axis-aligned
splitting planes, each defining the two child nodes of their parent
node. In order to keep the number of generated blocks small, the
splitting planes are chosen such that they minimize the number of intersections with the
bounding boxes of the subgrids in the domain represented by each node. 
Therefore we sweep the plane parallel to all three major coordinate
planes and determine the number of intersections. 
The split that introduces the smallest number of intersections
and has at least one slab of cells on each side, is chosen. In case
several such splits exist, we chose the one that divides the subgrids
in the most balanced way, in the sense that the ratio of the number of 
cells on each side is closest to $1$. The recursion stops, once a node 
covers only cells from the same subgrid.

\begin{figure}[t]
  \centering
     \begin{minipage}[c]{0.475\textwidth}
         \resizebox{1\columnwidth}{!}{ \includegraphics{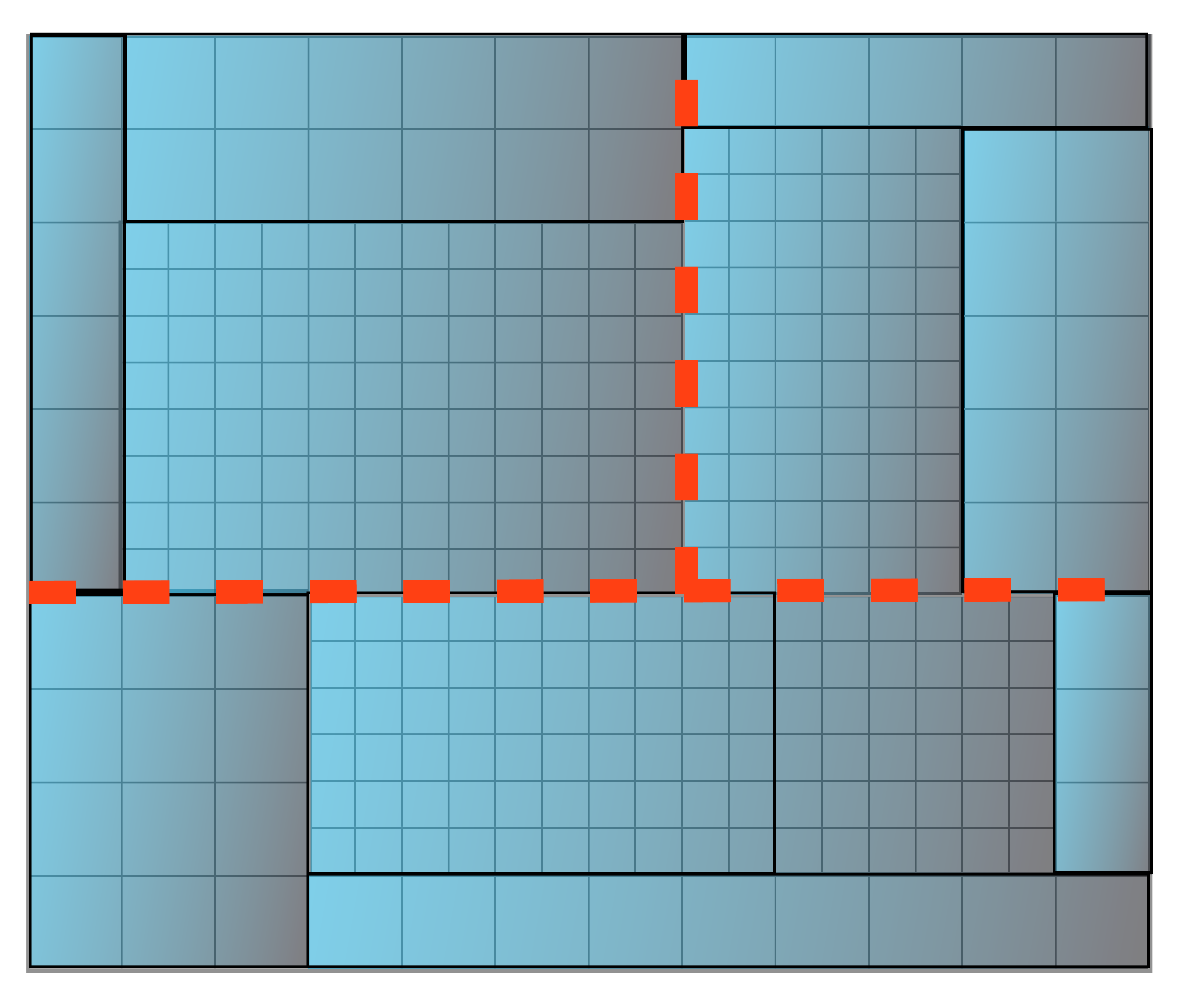} }
         \centerline{(a)}
     \end{minipage}
     \begin{minipage}[c]{0.475\textwidth}
        \resizebox{1\columnwidth}{!}{\includegraphics{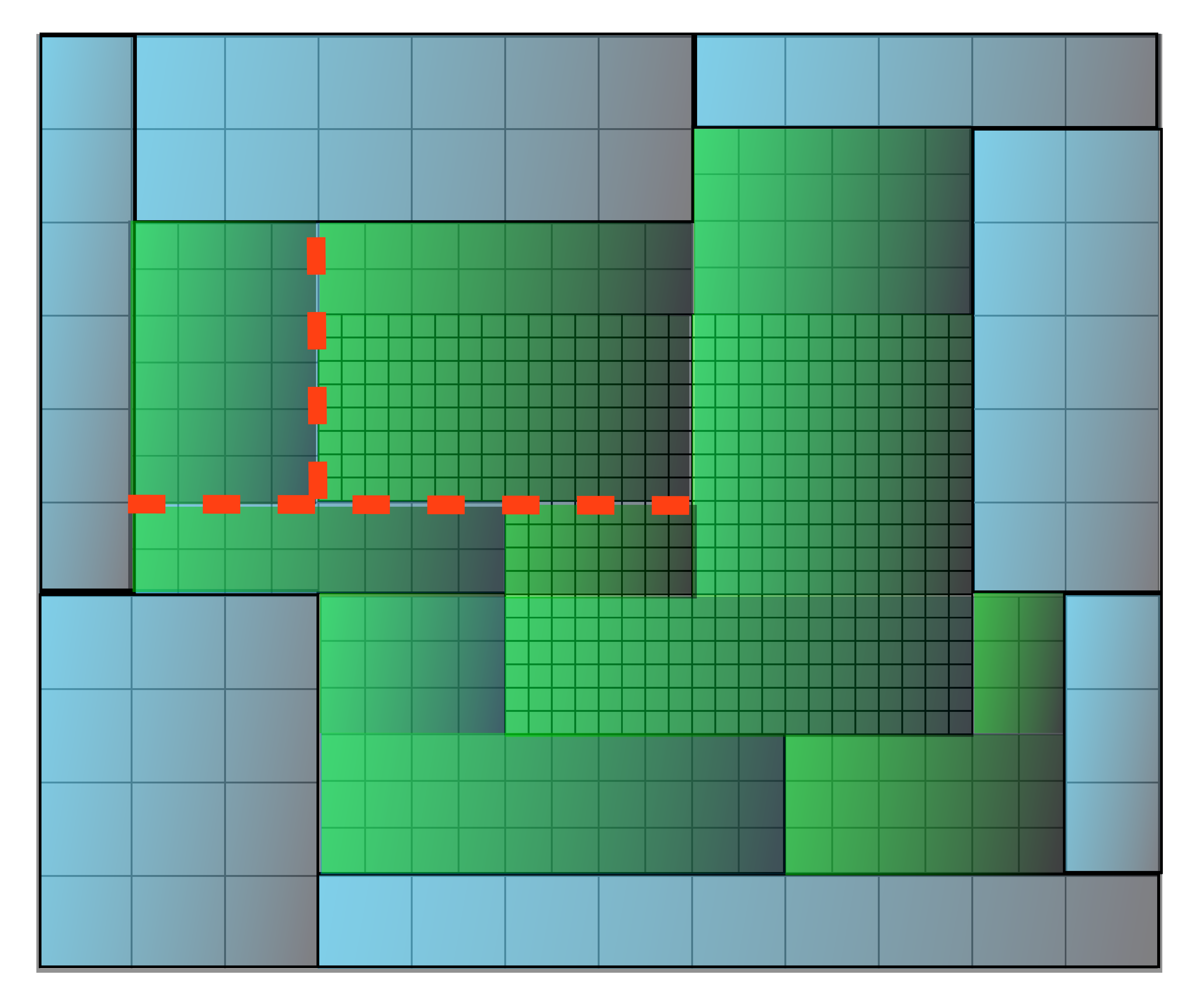} }
         \centerline{(b)}
     \end{minipage}
    \caption{Two-dimensional example of the decomposition procedure
      for the AMR hierarchy depicted in Figure~\ref{Fig::GridStructure}. Image (a) shows
    the resulting nodes of the kD-tree after the grids on the first
    level of refinement have been processed, image (b) shows the tree
    after all level 2 grids have been taken into account. To avoid
    cluttering only the first two splitting axis are depicted by the 
    red dotted lines. }
    \label{Fig:Kd-Tree}
\end{figure}

Next all subgrids $\Gamma^1_i$ of the first refinement level are
processed. For each leaf node in the current kD-tree we build a list with all the
subgrids $\Gamma^1_i$ that overlap with it. This can
be determined efficiently by traversing the current kD-tree top-down starting at the
root node, visiting only the child nodes that intersect
$\Gamma^1_i$. Next the kd-tree is
refined at each of the resulting leafs, by determining the optimal splitting planes for the blocks
defined by the intersections between the subgrids and the region
represented by the leaf node as discussed above. This
procedure is continued for the other refinement levels, successively
extending the tree for each level, until all subgrids $\Gamma^l_i$ in
the hierarchy are processed.
A 2D example for the subgrid configuration from Figure~\ref{Fig::GridStructure} is shown
in Figure~\ref{Fig:Kd-Tree}. The resulting kD-tree consists of three types of nodes: 
\begin{enumerate}[(a)]
  \item  nodes representing regions of the computational domain with
	cells that are further refined, 
  \item nodes that cover only cells that are unrefined, i.~e. leaf nodes of the kD-tree, and
  \item nodes that cover both, refined and unrefined cells, which are used to traverse 
       	the tree in a view-consistent order.
\end{enumerate}
The first type of nodes allows for a level-of-detail selection during
the rendering phase, as the corresponding region is covered by at
least two levels of resolution. If the resolution of the coarser level
is sufficient, which can for example be decided based on the projected
screen-space extension of the cells, the node is rendered at this
resolution and the traversal is stopped, otherwise the sub-tree of the
node is visited.
No data from the original AMR hierarchy is copied in this process, but
rather solely offsets and bounding box information as well as
references to the original subgrids are stored with the kD-tree.

\subsection{KD-TREE REPRESENTATION ON THE GPU}
\label{Sec::GPUkDtree}
To traverse the kD-tree structure on the GPU, we represent it by a set of 3D
textures. The first texture, called {\sl tree texture} in the following,
encodes the structure of the kD-tree, using one texel for each node.
Each texel consists of $64$~bits, split into $32$~bits for the {\sl red} and 
{\sl green} channel. The root node of the tree is stored at texel coordinates $(0,0,0)$.
The first two bits of the {\sl red}-channel encode the orientation 
of the splitting plane that defines the two child nodes and the next $6$~bits store the level of refinement of the corresponding block of cells. 
The remaining $24$~bits of the first channel are used to endcode the texel coordinates of
the first child node. The second child node is stored at the
next sequential texel.  As mentioned above the texel coordinate $(0,0,0)$ is
reserved for the root note, and we use it to indicate leaf nodes of the tree.

\begin{figure}[t]
  \centering
     \begin{minipage}[c]{0.95\textwidth}
         \resizebox{1\columnwidth}{!}{ \includegraphics{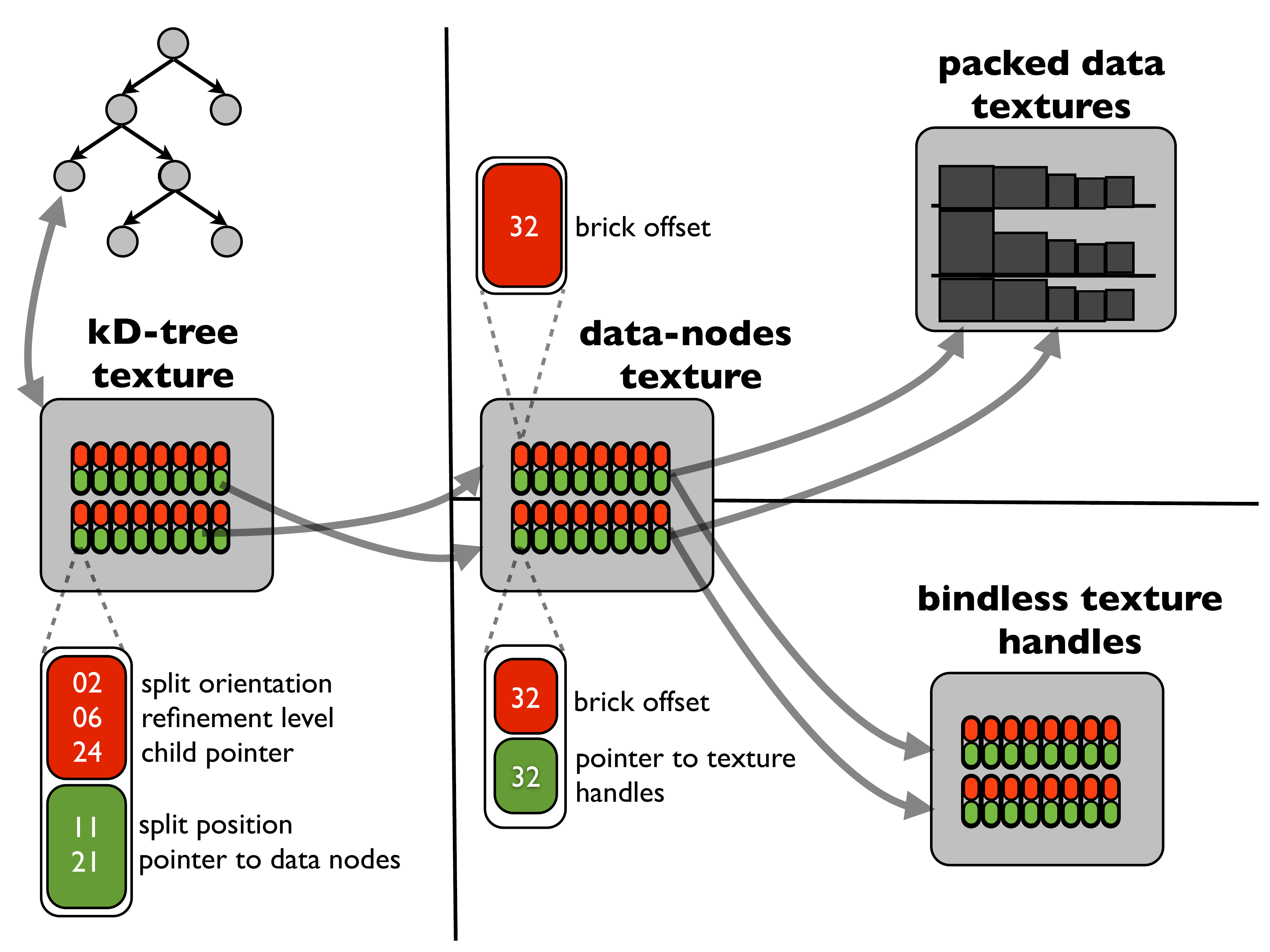} }
     \end{minipage}
    \caption{This figure shows the encoding of the kD-tree partition
      of the data domain using a set of 3D textures. The layout of
      the tree is stored using an integer valued 3D texture. Nodes of
      type (a) and (b), see Section~\ref{Sec::kDTree}, store indices into a
      second 3D texture that holds information for accessing the
      grid data associated with each node. The specific usage of its bits
      depends on the data storage strategy, see Section~\ref{Sec::GPUDataStorage}.}
    \label{Fig::GPU-REPRESENTATION}
\end{figure}

The first $11$ bits of the {\sl green}-channel store the location of the
splitting plane defining the child nodes. By construction of the tree,
the splitting planes are always located at the faces of the cells on the particular level of
refinement, so we do not need to store its value in floating point
coordinates. Instead it is beneficial to use integer coordinates,
defined as the number of cells on the current level of refinement, relative to the
node's lower left corner. The remaining $21$~bits of the second channel 
are used as an index into a second 3D-texture that holds specific
information required for nodes of type (a) and (b), see
Section~\ref{Sec::kDTree}, and will be discussed in the next subsection.
A diagram that depicts the specific usage of bits is shown in 
Figure~\ref{Fig::GPU-REPRESENTATION}.

A $256^3$ index texture, with a memory requirement of $128$
MBytes, is capable of encoding kD-trees with more than $16$
million nodes. The specific choice of bits allows us to distinguish $64$ levels of
refinement and subdivision plane positions for nodes covering up to
$2048^3$ cells on their level of refinement, sufficient for the
largest AMR simulations up to date. 
For moderately sized AMR hierarchies usually a resolution for $128^3$
index texture, with memory requirements of only $16$ Mbytes, enabling
the storage of more than $2$ million nodes, is sufficient.

\section{DATA STORAGE ON THE GPU}
\label{Sec::GPUDataStorage}
As discussed in Section~\ref{Sec::GPUkDtree}, texels for 
nodes of type (a) or (b), i.e.~nodes that represent block of cells
that are either completely refined or completely unrefined and thus
can be rendered,
store an index into a second texture, 
called {\sl data nodes} texture in the following. The specific usage of
its bits depends on the storage strategy for the data associated with the AMR grids. 
One challenge for GPU-raycasting of multi-resolution data is that
only a limited number of textures can be accessed simultaneously,
depending on the number of texture units of the specific graphics
hardware. The maximal number is currently about $100$ , far too few
to assign a separate texture to each node in the tree structure.
Typical AMR simulations generate between $10^3$ to $10^5$ separate
subgrids for each time step.
One option to tackle this problem is to use a large 3D texture as a
memory pool and copy the data blocks associated with each AMR grid into
this texture, which will be discussed in Section~\ref{SubSec::MemoryPool}.
In Subsection~\ref{SubSec::BindlessTextures} we will discuss an
alternative approach, based on {\sl NVIDIA's Bindless
 Textures} extension for {\sl OpenGL},~\cite{BindlessTexturesWebsite}
available since March 2012, which enables {\sl OpenGL} applications
to dynamically access large number of separate textures in the
graphics shaders.

In both cases it is advantageous to assign a separate texture brick per
subgrid instead of one brick for each kD-tree nodes of type (a) and
(b), because in general there are more kD-tree nodes than
subgrids and for higher order interpolation we use a common row of 
texels at interfaces between adjacent texture blocks. Assigning
a separate texture brick per kD-node would drastically increase the 
number of interfaces and thus texture memory consumption. Furthermore 
the packing procedure would result in more fragmented areas for a
larger number of smaller bricks, see
Section~\ref{SubSec::MemoryPool} and~\ref{Sec::Results}.
So we allocate one brick for each AMR subgrid and rather store
offsets into these bricks at the nodes of type (a) and (b).
We employ nearest-neighbor interpolation for cell-centered 
AMR data and trilinear interpolation for vertex-centered data. In the first 
case the texels are aligned with the centers of the cells, while in the second one they
are aligned with the vertices of the grid. To avoid artifacts originating from 
discontinuous tri-linear interpolation between subgrids with different
resolutions, adjacent 
texture-blocks share a row of data samples at their common boundary faces and
the data at dangling nodes has to be replaced to the interpolated texel values 
of the abutting, coarse texture.

\subsubsection{TEXTURE PACKING APPROACH}
\label{SubSec::MemoryPool}
For our purposes the following variant of the three-dimensional packing 
    problem is appropriate: pack a given number of
    axis-aligned rectilinear boxes into one container with fixed width
    and depth, such that its height is minimized.~\cite{li90threedimensional}
    This problem belongs to the class of NP-hard problems, but a couple of 
    useful heuristics have been suggested.
    Similar to the approach discussed in Kaehler~et~al.~\cite{Kaehler01sparseVolren} we
    use the so-called {\sl next-fit-decreasing-height (NFDH)}
    algorithm.~\cite{cof:1980:joh}
    First the texture bricks are inserted into a list, in the order of
    decreasing extension in the z, y and then x-direction. 
    The packing algorithm starts at the lower left-hand corner of the container 
    and inserts the boxes along the x-axis until the maximal x-extension of the container is reached.
    A new row is opened, with a y-coordinate given by the largest y-extension 
    of the already inserted boxes. This procedure is repeated until the lowest 
    layer of the container is filled. Then a new layer is in the
    z-direction is opened and this process  continues until all boxes are inserted. 
    We iterate this procedure with different values for the base layer
    extensions of the container and the result with the smallest
    volume is chosen. A 3D-texture of this size is defined with 
    the subtextures inserted at their computed positions. 
   Each kD-node of type (a) or (b) in the {\sl ``data nodes texture''}, see upper-right part of
   Figure~\ref{Fig::GPU-REPRESENTATION}, stores its offset into the packed
   texture using 32-bits. This allows to index into a
   packed texture of up to $2048 \times 2048 \times 1048$ texels.

\begin{figure}[t]
  \centering
     \begin{minipage}[c]{0.6\textwidth}
        \resizebox{1\columnwidth}{!}{\includegraphics{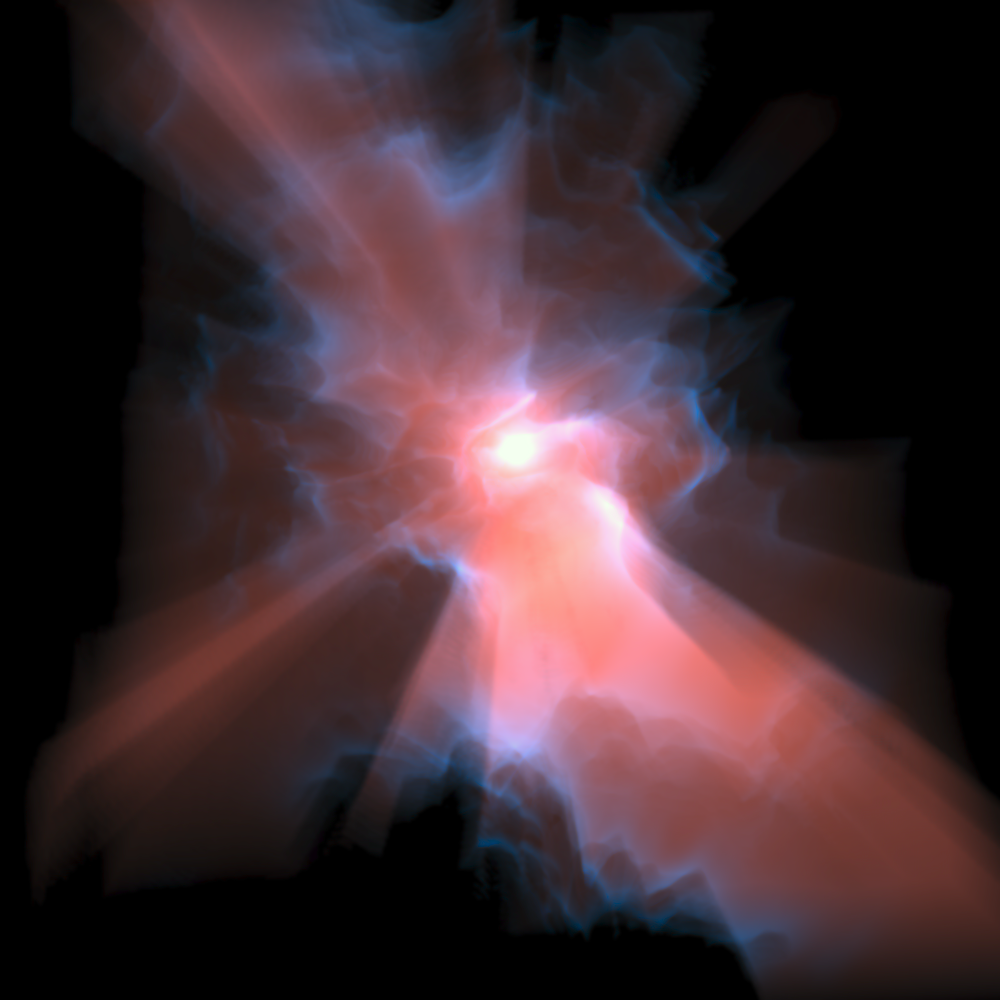} }
    \end{minipage}
    \caption{A global illumination example, tracing secondary rays to 
      a central point-light source that illuminates the whole domain. }
    \label{Fig:AdvancedRendering2}
\end{figure}

\begin{figure}[t]
  \centering
     \begin{minipage}[c]{0.6\textwidth}
         \resizebox{1\columnwidth}{!}{ \includegraphics{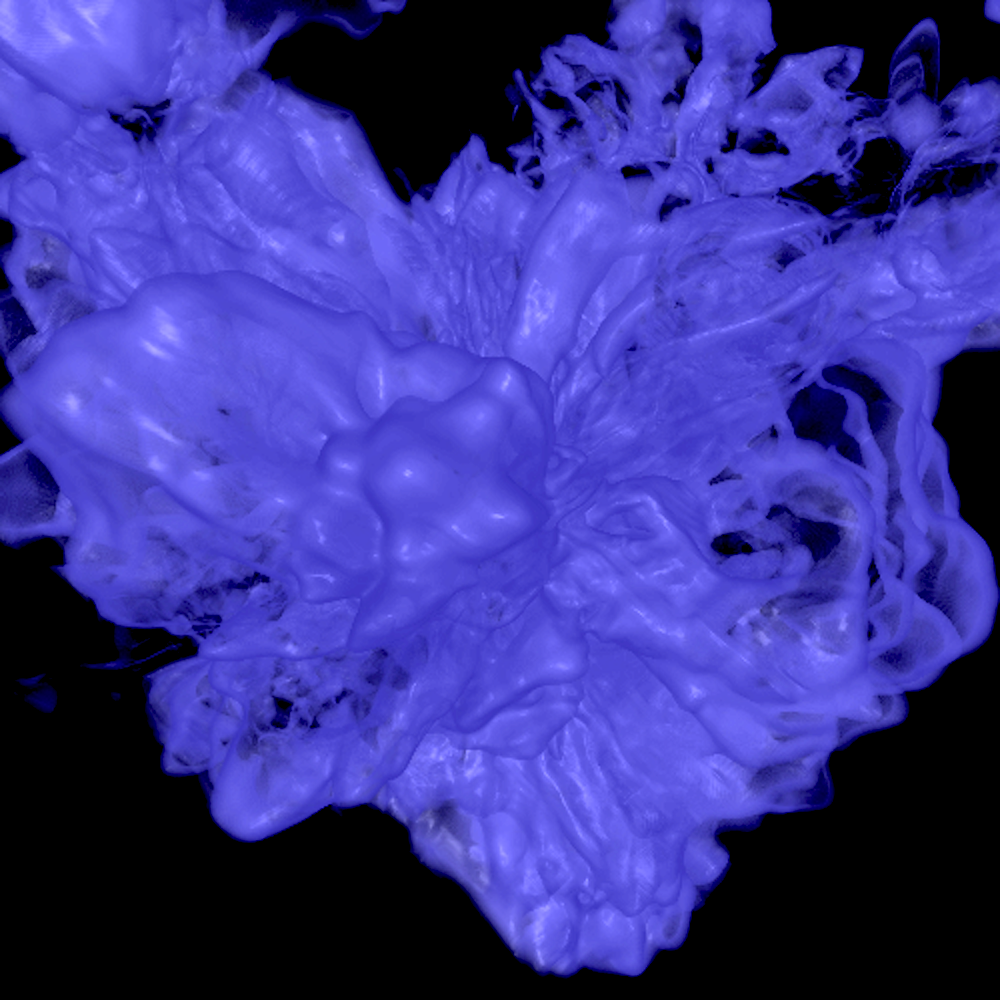}}
    \end{minipage}
  \caption{A non-polygonal, semi-transparent
      iso-surface representation using a gradient-based shading
      approach, with {\sl on-the-fly} gradient computation.}
   \label{Fig:AdvancedRendering1}
\end{figure}

\subsubsection{DATA ACCESS USING THE BINDLESS TEXTURES EXTENSION}
\label{SubSec::BindlessTextures}
{\sl NVIDIA}'s {\sl Bindless Texture}
extension~\cite{BindlessTexturesWebsite} allows
{\sl OpenGL} applications to dynamically access large numbers of texture objects
in graphics shaders without the need to first bind the textures to specific texture units on the CPU.
Instead each texture is identified by a 64-bit handle that is used
to sample the texture. This provides a means to manage the
large amounts of separate texture bricks associated with typical AMR data
structures without the need to pack them into a memory pool.
We use 32-bits per texel for the {\sl data nodes} texture in this case.
For each kD-tree node of type (a) and (b) we employ a 32-bit index 
into another 3D-texture with two 32-bit channels, referred to 
as the {\sl handles}  textures in the following. It endcodes the $64$-bit
texture handles for each texture brick associated with a subgrid, 
see Figure~\ref{Fig::GPU-REPRESENTATION}. 
It is advantageous to store the handles in a separate texture
because the number of subgrids is much smaller than the
number of kD-nodes, so storing them at each entry in the {\sl data
nodes} texture would introduce an overhead in GPU-memory usage.

\subsection{RAY TRAVERSAL}
\label{Sec::Traversal}
As in the standard GPU-raycasting approach for uniform data~\cite{stegmaier},
we draw the faces of the bounding box enclosing the computational
domain to execute an instance of a fragment shader for each covered pixel.
In the fragment shader the ray's origin and direction for the corresponding pixel is computed.
Next the segments resulting from the intersection between the
viewing ray and the kD-tree data nodes are determined similar to the
{\sl kD-restart} algorithm~\cite{Horn:2007:IKT:1230100.1230129}.
 The {\sl kd-tree} texture is sampled starting at the root
node with texel coordinate $(0,0,0)$ and traversed
{\sl top-down}, using the child node pointers stored at each node,
as discussed in Section~\ref{Sec::GPUkDtree}.  The bounding box of
each node is computed on-the-fly from the extensions of
the kD-root node and the orientation and
positions of the splitting planes for each node. The
split position is mapped from integer-coordinates to
the world coordinate system, using
the cell size ${\bf h}^0$ on the root level, the current level of
refinement at the node as well as its bounding box. The traversal continues
until either a leaf node is reached, indicated by an ``invalid''
child node entry of $(0,0,0)$,  or until a node of type (a)
is visited, which is the case if the node has a valid child node
entry and an index into the {\sl data nodes} texture. The latter case
allows for a level-of-detail selection, pruning the traversal of the tree
if for example the projected screen size of the cells of this
node is below a user-defined threshold.

Next the color and opacity contribution of the corresponding ray-segment is
computed. In case of the ``packing'' approach discussed in
Section~\ref{SubSec::MemoryPool}, 
the node's offset into the packed texture is sampled
from its entry in the  {\sl data nodes} texture and the ray-segment is
transformed to texel coordinates.

In the {\sl Bindless  Texture} approach the
current texture handles are read from the  {\sl handles} texture
using the index stored in the {\sl data nodes} texture, and 
converted to a {\sl GLSL sampler3D} object. The ray-position is
converted to texture coordinates using the number of samples of the
texture and the number of texels of the subregion corresponding to
the kD-node, which can be computed 
on-the-fly from its bounding box and the current refinement level.
The sampling rate is chosen proportional to the level's cell-size.
When the segment is processed, the kD-tree traversal is ``restarted''
at the root node and the next ray-segment is visited.  Once the total
ray is processed, the resulting colors and opacities are written to the frame-buffer.

\section{Results and Discussion} 
\label{Sec::Results}  
The comparison was performed using a {\sl NVIDIA GeForce GTX 680} 
graphics card with $2$ GByte of graphics memory, that was installed on a host with a 
 {\sl Intel Xeon E5520 CPU} and $24$ GByte main memory. The rendering
 algorithms were implemented in {\sl OpenGL} and the {\sl OpenGL Shading Language (GLSL)}. 
We tested the performance and memory requirements of the proposed
algorithms on three datasets with different sizes and characteristics.
All performance measurements refer to a viewport size of $1000^2$ pixels.
Table~\ref{Tab::Datasets} lists information about the
datasets and the corresponding kD-trees: the number of subgrids, 
refinement levels and cells in the original SAMR
grid hierarchies as well as the total number of nodes in the resulting 
kD-trees and the portion of nodes of type (b) and (c),
see Section~\ref{Sec::kDTree}.

\begin{table}[h]
\centering
\caption{The characteristics of the datasets: the
  number of subgrids, refinement levels, cells as well as the total
  number of nodes in the resulting kD-trees and the portion of
  internal and leaf nodes that are associated with blocks of cells. }
\label{Tab::Datasets}

\begin{tabular}{|r||r|r|r|r|r|r|}
\hline
 & \#grids & \#levels & \#cells & \#kd-tree nodes & \#kd-data nodes\\
\hline 
dataset 1 & 2,666 & 4 & $19 \times 10^6$ & 30,096 &  27,249\\
dataset 2 &  18,528 & 4  & $33 \times 10^6$ & 178,114 &  162,095 \\ 
dataset 3 & 39,061  & 13 & $140 \times 10^6$ & 368,225 & 343,389 \\
\hline
\end{tabular}

\end{table}

\begin{figure}[H]
  \centering
     \begin{minipage}[c]{0.3\textwidth}
         \resizebox{1\columnwidth}{!}{ \includegraphics{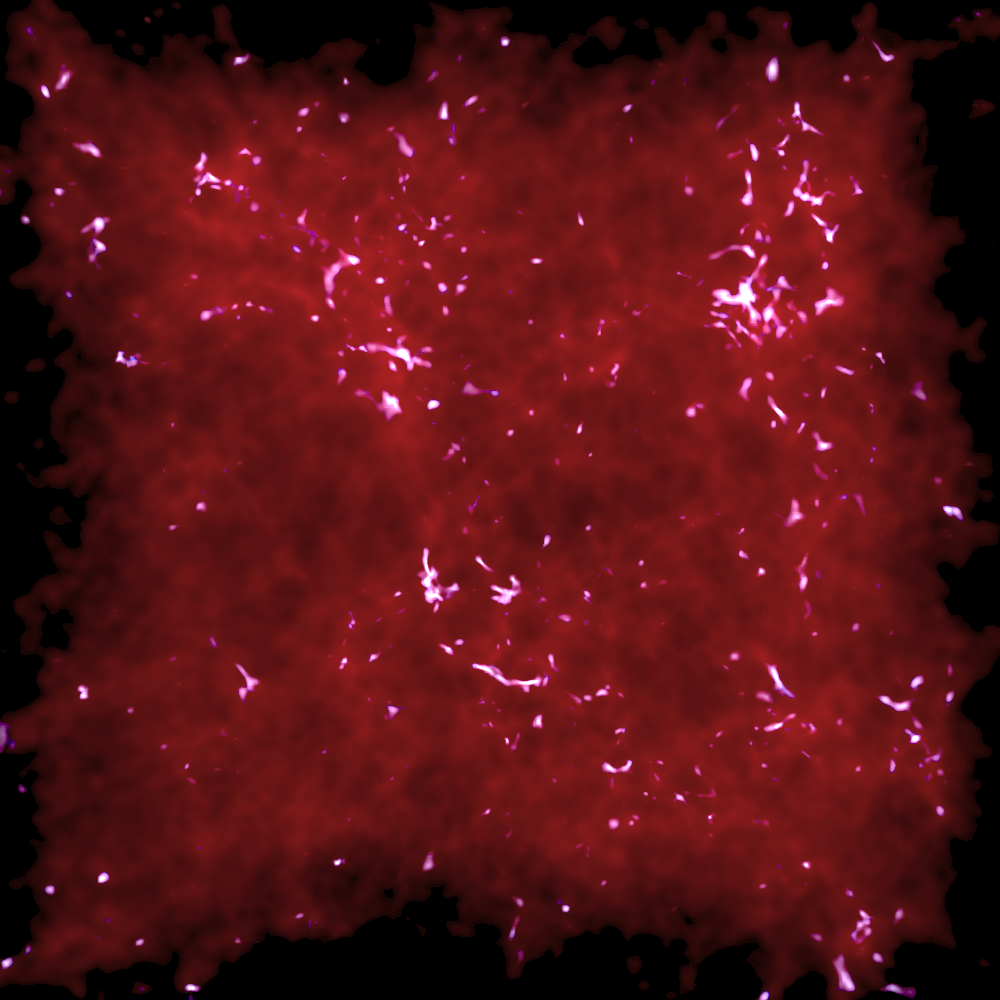} }
         \centerline{(a)}
         \vspace{0.0001cm}
     \end{minipage}
     \begin{minipage}[c]{0.3\textwidth}
        \resizebox{1\columnwidth}{!}{\includegraphics{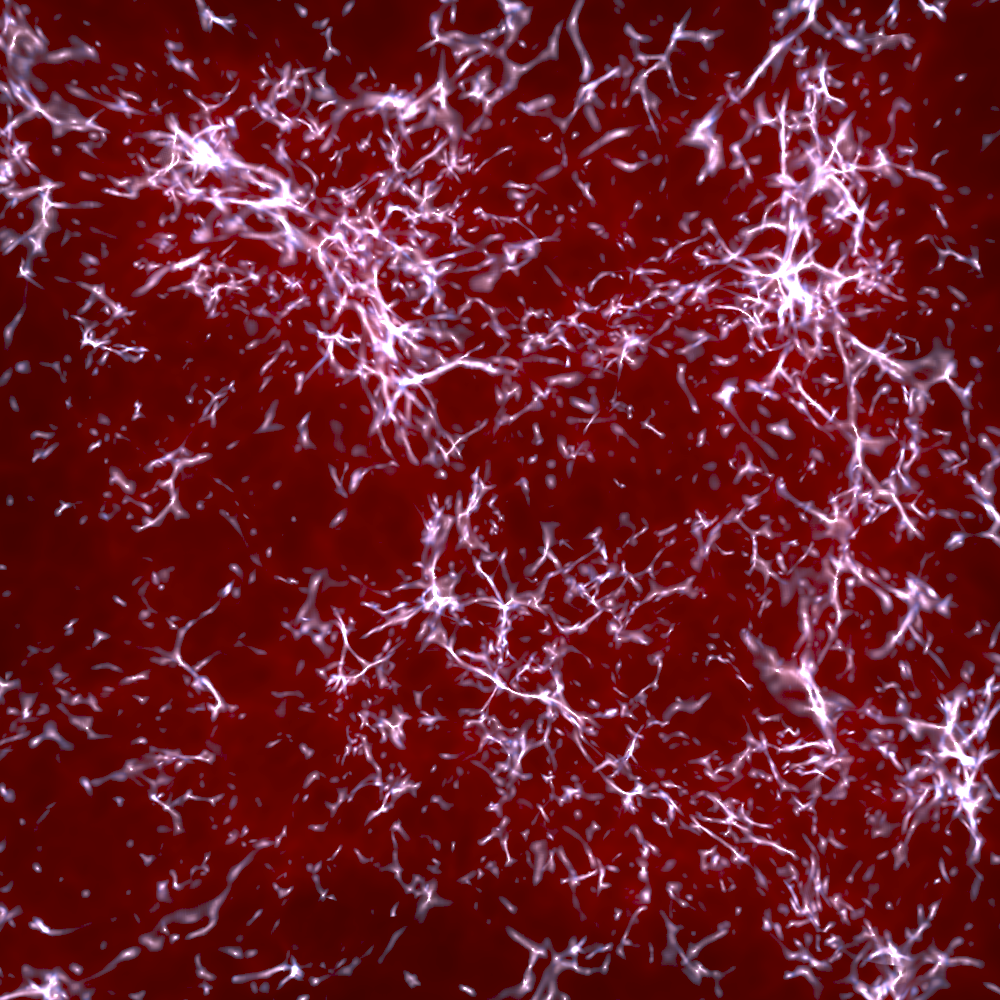} }
         \centerline{(b)}
     \vspace{0.0001cm}
     \end{minipage}
    \begin{minipage}[c]{0.6\textwidth}
        \resizebox{1\columnwidth}{!}{\includegraphics{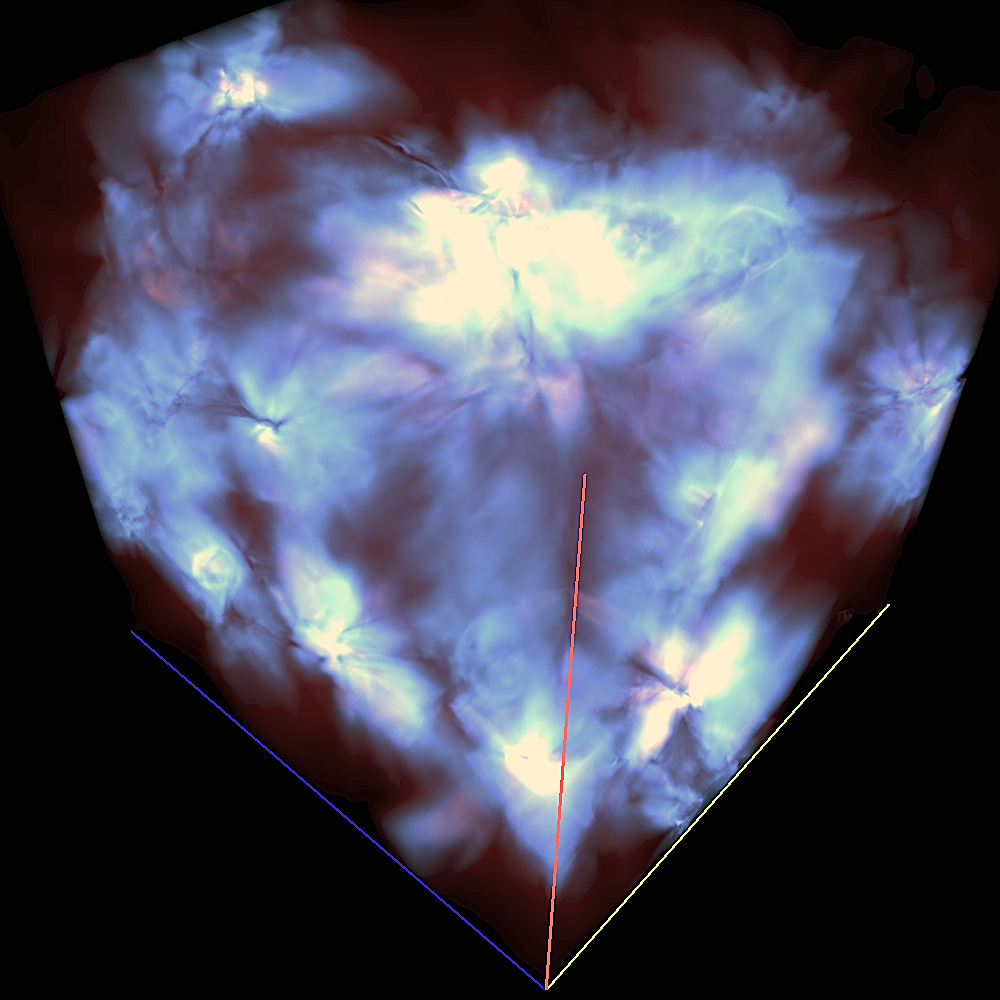} }
         \centerline{(c)}
     \end{minipage}
    \caption{Renderings of the three different data sets used in
    this paper. Image (a) and (b) show the large scale distribution of
    hydrogen gas on scales of 100 Mpc at two different time
    steps. Image (c) shows the temperature distribution between
    dwarf galaxies that formed in the early Universe.
    Information about the datasets can be found in Table~\ref{Tab::Datasets}. }
    \label{Fig:Datasets}
\end{figure}

We compared the two single-pass rendering methods proposed in this
paper to a  multi-pass approach~\cite{vg06-kaehler},  that traverses the kD-tree on the
CPU and renders each data node separately, by first binding the associated
texture to a texture unit and rendering the bounding box of the
kD-tree node to initialize the fragment shaders. The kD-tree 
partition approach described in Section~\ref{Sec::kDTree} was used for all three methods.
Table~\ref{Tab::Timings} shows the GPU memory requirements, preprocessing times and performance
numbers for the different rendering methods. The numbers in
each cell of the table are the measurements for the three different datasets.
For the {\sl packing} and the {\sl bindless} textures approach the first number in the ``GPU
memory'' column is the size of the set of the 3D integer textures used to encode the
kD-tree structure and the data nodes, whereas the second number gives the memory requirements
for the grid data, i.~e. the packed texture or the sum of the separate 3D-textures 
in the {\sl bindless} case. An emission-absorption model with no
further acceleration techniques, like {\sl early-ray-termination} or
{\sl empty-space-skipping}, was used in the examples. Renderings of the different
datasets are shown in Figure~\ref{Fig:Datasets}.

\begin{table}[h]
\centering
\caption{GPU memory requirements, preprocessing times and performance
numbers for the three different rendering methods. The numbers in
each table cell are measurements for the three different datasets
shown in Table~\ref{Tab::Datasets} and Figure~\ref{Fig:Datasets}. }
\label{Tab::Timings}
\begin{tabular}{|r||r|r|r||r|r|r||r|r|r||}
\hline
 & \multicolumn{3}{|c||}{GPU memory [Mbytes]} 
 & \multicolumn{3}{|c||}{preprocessing[s]} 
 & \multicolumn{3}{|c||}{performance [fps]}\\
\hline
multi-pass           & 71.4            & 124.2           & 530.3 & 1.1 & 2.6 & 6.8 & 4.2 & 2.1 & 0.8 \\
\hline
packing                & 0.5 + 239.3 & 2.7 + 310.3 & 4.3 + 1000.2 & 3.4 & 7.5 & 8.1 & 3.2 & 1.6 & 1.2 \\ 
\hline
bindless-texture  & 0.5 + 71.4  & 2.8 + 124.2 & 6.1 + 530.3  & 1.7 & 3.2 & 7.4  & 2.1 & 0.9 & 0.4 \\
\hline
\end{tabular}
\end{table}

As indicated by the measurements shown in Table~\ref{Tab::Timings} the
rendering performance of the packing approach is faster than the {\sl bindless
texture} approach for all examples, due to the overhead associated with the
dynamic access of the separate textures in the {\sl bindless
texture} case. However, the packing approach uses more texture
memory, as the packing of the differently sized subgrid textures 
into the texture memory pool necessarily introduces some
fragmentation. An efficiency, defined as the number of used
texels to the total number of texels in the packed texture, between 30\% and 50\% was achieved  
for the three datasets. The multi-pass approach has faster rendering performance for
the smallest and the medium sized dataset 1 and 2, but the packing approach is
about 50\% faster for the largest dataset, number 3. Here the cost of
the per-pixel sampling of the kD-tree structure is lower than the
overhead for binding of the separate textures and rendering the
bounding boxes of each kD-node to initialize the fragment shader 
instances in the multi-pass approach. 

Unlike the multipass approach the single-pass methods allow the application of 
advanced lighting schemes, that require simultaneous access to more
than one kD-tree node. Two examples for dataset~3 
are shown in Figure~\ref{Fig:AdvancedRendering2} and~\ref{Fig:AdvancedRendering1}. 
Figure~\ref{Fig:AdvancedRendering1} is a non-polygonal
semi-transparent iso-surface representation using a gradient-based shading
approach, rendered at $0.8$ fps. The gradients were computed 
{\sl on-the-fly}. Figure~\ref{Fig:AdvancedRendering2} shows a global illumination example. Here for
each sampling location a secondary ray is traced to a central
point-like light source. The achieved frame rate was $0.1$ fps.

\section{Conclusions}
\label{Sec::Conclusions}
We presented a single-pass GPU-raycasting approach for
{\sl Structured Adaptive Mesh Refinement (SAMR)} data. It employs a
kD-tree to subdivide the data domain into axis-aligned, non-overlapping blocks of cells 
from the same resolution level. The tree is encoded by a set of 
3D-textures, which allows to efficiently traverse it entirely
on the GPU. We discussed two different data access strategies,
namely a ``packing'' approach using a texture memory pool, 
and a method based on {\sl NVIDIA's Bindless Texture}
extension for {\sl OpenGL}, and applied them to several SAMR datasets 
of different sizes and complexity.

For all examples the 3D textures used to encode the
kD-tree structure required only small amounts of texture memory. 
The packing approach offers higher rendering performance as long as
all data fits into texture memory, because of the 
extra costs of the dynamically accessing the separate textures in the 
{\sl  Bindless Texture} approach, whereas the latter consumes less texture memory.
We further compared the new approaches to a  previously published
multi-pass method~\cite{vg06-kaehler}. For complex and large SAMR
datasets the packing approach outperformed the multi-pass
algorithm and in contrast to the latter, both new methods enable a
straight-forward implementation of many advanced shading and acceleration techniques,
since all parts of the data domain are accessible in the fragment
shader. They further do not suffer from {\sl read-after-write hazards} 
as multi-pass approaches that use non-standard blending 
equations and need to read back from the frame-buffer for each pass,
or apply synchronization methods, which substantially decrease the rendering performance.
Because the new single-pass methods are executed entirely on the GPU without
any CPU interaction, their rendering performance should directly benefit
from the increased number of shader cores expected for upcoming GPU
generations. The {\sl bindless texture} approach is especially well
suited for datasets that exceed the available graphics memory, as it
allows to dynamically upload subsets of textures as required by out-of-core
rendering approaches.

%%-----------------------------------------------------------
\section{Acknowledgments} 
We would like to thank Ziba Mahdavi (KIPAC) and Shalini
Venkataraman (NVIDIA) for providing us with $NVIDIA$ $Kepler$ GPUs.
We also thank Ji-hoon Kim (UCSC) and John Wise (Georgia Tech) for the
datasets used to test the presented algorithms.
This work was supported in part by the {\sl National Science Foundation} 
through award number AST-0808398 and the LDRD program at the 
SLAC National Accelerator Laboratory as well as the Terman fellowship 
at Stanford University. 

%%%%%%%%%%%%%%%%%%%%%%%%%%%%%%%%%%%%%%%%%%%%%%%%%%%%%%%%%%%%%
%%%%% References %%%%%

%%\bibliography{ralf}   %>>>> bibliography data in report.bib
%\include{references.bbl}
\bibliographystyle{spiebib}   %>>>> makes bibtex use spiebib.bst

\end{document}